\documentclass[
reprint,
 amsmath,amssymb,
 aps,
 prapplied,
longbibliography]{revtex4-2}

\usepackage{graphicx}
\usepackage{dcolumn}
\usepackage{bm}

\usepackage[english]{babel}

\usepackage{natbib}

\usepackage{datetime2}

\usepackage{letltxmacro}

\LetLtxMacro{\ORIGselectlanguage}{\selectlanguage}
\makeatletter
\DeclareRobustCommand{\selectlanguage}[1]{%
  \@ifundefined{alias@\string#1}
    {\ORIGselectlanguage{#1}}
    {\begingroup\edef\x{\endgroup
       \noexpand\ORIGselectlanguage{\@nameuse{alias@#1}}}\x}%
}
\newcommand{\definelanguagealias}[2]{%
  \@namedef{alias@#1}{#2}%
}
\makeatother

\definelanguagealias{en}{english}

\usepackage{amssymb}

\usepackage{comment}

\usepackage{xcolor}

\usepackage{graphicx}
\usepackage{dcolumn}
\usepackage{bm}

\usepackage[utf8]{inputenc}
\usepackage[T1]{fontenc}
\usepackage{mathptmx}
\usepackage{etoolbox}

\makeatletter
\def\@email#1#2{%
 \endgroup
 \patchcmd{\titleblock@produce}
  {\frontmatter@RRAPformat}
  {\frontmatter@RRAPformat{\produce@RRAP{*#1\href{mailto:#2}{#2}}}\frontmatter@RRAPformat}
  {}{}
}%
\makeatother

\begin{document}

\title{A Zero-Bias Superconducting Voltage Amplifier Based on \\
the Bipolar Thermoelectric Effect}
\author{G. Trupiano}
\email[Corresponding author: ]{giacomo.trupiano@sns.it}
 \affiliation{NEST, Istituto Nanoscienze-CNR and Scuola Normale Superiore, Piazza S. Silvestro 12, I-56127 Pisa, Italy}
 
\author{G. De Simoni}
\affiliation{NEST, Istituto Nanoscienze-CNR and Scuola Normale Superiore, Piazza S. Silvestro 12, I-56127 Pisa, Italy}

\author{F. Giazotto}
    \affiliation{NEST, Istituto Nanoscienze-CNR and Scuola Normale Superiore, Piazza S. Silvestro 12, I-56127 Pisa, Italy}


\begin{abstract}

We introduce a zero-bias superconducting voltage amplifier that harvests energy from a thermal gradient by exploiting negative differential resistance (NDR) in an asymmetric tunnel junction. The device is based on an asymmetric superconductor–insulator–superconductor (SIS) junction with an energy-gap ratio of $\Delta_1/\Delta_2 = 0.5$, connected in series with a load resistor. Owing to the superconducting bipolar thermoelectric effect, the current–voltage characteristic of the junction exhibits a region of NDR, in which the net current flows opposite to the applied voltage. This mechanism enables voltage amplification in the absence of any external electrical bias, relying solely on the temperature difference between the electrodes ($T_H \simeq 1$~K, $T_B \simeq 20$~mK). 
Numerical simulations predict a voltage gain of 20~dB, a 1~dB compression point at an input amplitude of 2~$\mu$V, and a total harmonic distortion below $-50$~dB. The input-referred noise is approximately $1\;\text{nV}/\sqrt{\text{Hz}}$, with an associated thermal load on the order of nanowatts. The frequency response is broadband from near DC, with a $-3$~dB cutoff around 180~MHz, set by the RC time constant of the junction. Using Al-, Al-Cu-, and AlO$_x$-based technologies, the amplifier is compatible with conventional superconducting circuit fabrication processes. These findings demonstrate that thermoelectric superconducting junctions can deliver bias-free voltage amplification from near DC up to $\sim 200$~MHz, making them promising candidates for transition-edge sensor readout, quantum circuit instrumentation, and low-frequency cryogenic signal processing.

\end{abstract}


\maketitle

\section{Introduction} \label{sec:introduction}

\begin{figure*}[t!]
\includegraphics[width=\linewidth]{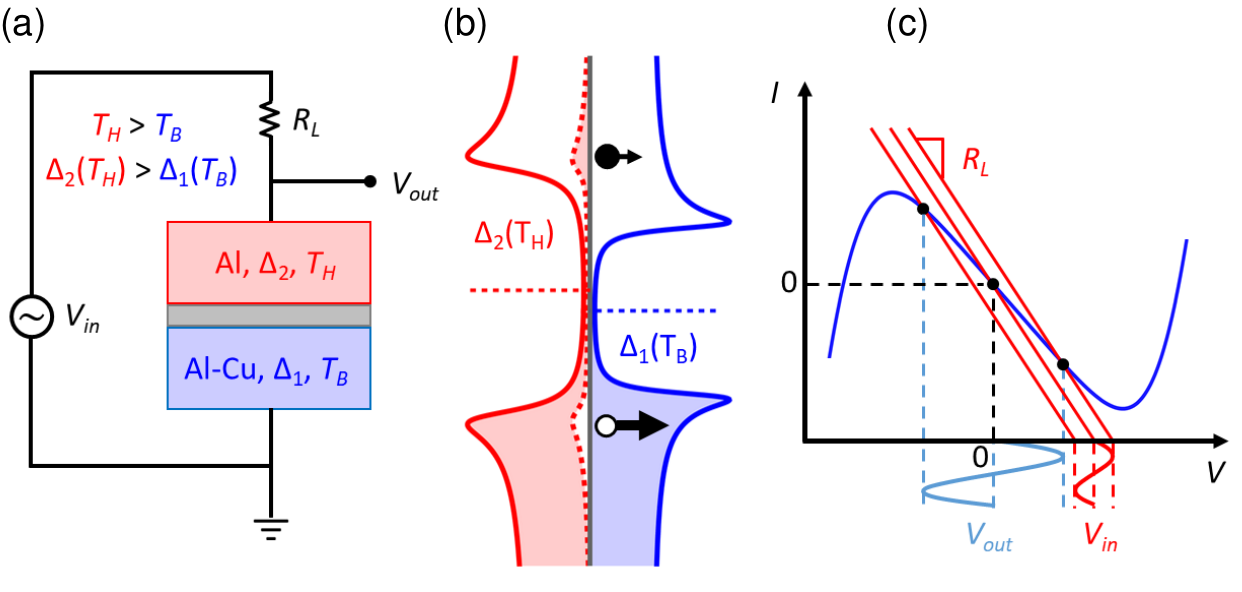}

\caption{\label{fig:fig1} Device concept and operating principle.  
(a) Device schematic. The amplifier comprises an asymmetric superconductor–insulator–superconductor (SIS) tunnel junction ($\text{Al}/\text{AlO}_x/\text{Al-Cu}$) connected in series with a load resistor $R_L$. The superconducting energy gaps of the two electrodes are $\Delta_2(T=0) = 0.20$~meV and $\Delta_1(T=0) = 0.10$~meV, respectively. A thermal gradient ($T_H \simeq 1$~K, $T_B = 20$~mK) provides the driving energy, such that the device operates as a bias-free voltage divider.  
(b) Energy-band diagram illustrating the superconducting bipolar thermoelectric effect. The combination of gap asymmetry ($\Delta_1/\Delta_2 = 0.5$) and thermal gradient ($T_H \gg T_B$) ensures the condition $\Delta_2(T_H) > \Delta_1(T_B)$. Thermal broadening of the quasiparticle distribution in the hot electrode leads to an effective alignment of quasiparticle energies with the gap-edge density-of-states (DOS) singularities in the cold electrode. This specific energy alignment induces a net charge current flowing opposite to the applied voltage for $|V| \lesssim \Delta_2(T_H) - \Delta_1(T_B)$, thereby yielding a negative absolute resistance ($IV < 0$) and a negative differential resistance $r_d = (\text{d}I/\text{d}V)^{-1} < 0$ at zero applied bias.  
(c) Voltage amplification mechanism. The current–voltage ($I$–$V$) characteristic of the junction (blue curve) intersects the load line determined by $R_L$ (red line) at a zero-bias operating point. A small AC input signal $V_{in}$ modulates this operating point, giving rise to an amplified output voltage $V_{out}$. The corresponding voltage gain is given by $G_V = |r_d|/(|r_d| - R_L)$, with $G_V > 1$ for $R_L < |r_d|$. The amplification process is fully powered by the thermal gradient; no external DC power is required.}
\label{fig:device_concept}
\end{figure*}

Cryogenic amplification is a critical component of signal readout for quantum devices, superconducting detectors, and nanoscale sensors operating at millikelvin temperatures. Existing superconducting amplifier technologies are predominantly based on flux-sensitive or current-biased architectures, each optimized for distinct frequency regimes. Superconducting quantum interference device (SQUID) amplifiers offer low-noise gain for low-frequency signals. However, they function intrinsically as flux-to-voltage transducers and consequently require magnetic coupling to the input signal~\cite{clarke2004handbook}. At microwave frequencies, Josephson parametric amplifiers (JPAs)~\cite{castellanos2007widely} and traveling-wave parametric amplifiers (TWPAs)~\cite{macklin2015near} can realize noise performance approaching the quantum limit, yet are typically constrained to relatively narrow operational bandwidths and depend on current or flux coupling to the input.

In contrast, superconducting amplifiers that provide direct voltage input and output over the DC-100 MHz range remain comparatively underexplored. This frequency band is technologically relevant for a variety of applications, including multiplexed cryogenic detector readout~\cite{irwin2005transition}, transition-edge sensors~\cite{irwin2005transition,de2024transition}, superconducting single-photon detectors~\cite{McCaughan2018}, and nanomechanical resonators~\cite{Ekinci2005}. 
Low-frequency cryogenic amplification in these contexts is conventionally realized using semiconductor-based electronics. Such amplifiers typically exhibit input-referred voltage noise on the order of $1\;\text{nV}/\sqrt{\text{Hz}}$ with bandwidths extending from DC to several megahertz, but they dissipate power at the milliwatt level. As a result, they are generally placed in the $4\,\mathrm{K}$ stage of the dilution refrigerators rather than in the millikelvin stage~\cite{Woods2015,Zavjalov2019,Sidorenko2020, 8036394}. This architecture increases wiring complexity and introduces additional parasitic capacitance between the device under test and the first amplification stage. 
A superconducting voltage amplifier capable of operating directly at millikelvin temperatures, with nanowatt-level power dissipation, would enable fully integrated cryogenic instrumentation while preserving low-noise performance.

Voltage amplification based on negative differential resistance (NDR) is a well-established principle in semiconductor device physics. A prototypical implementation is the resonant tunneling diode (RTD)~\cite{chang1974resonant}, in which quantum-mechanical resonant tunneling through a double-barrier heterostructure gives rise to an NDR region in the current-voltage characteristic ($I$–$V$), thus allowing voltage gain in an appropriate voltage-divider configuration. However, RTDs require a finite direct-current (DC) bias to operate in the NDR regime, resulting in static power dissipation and rendering them unsuitable for integration at the millikelvin stage. 
The device introduced in this work leverages an analogous NDR mechanism realized in a superconducting tunnel junction, with two crucial distinctions: first, the NDR region is centered at zero applied bias, such that no DC power is dissipated at the operating point; second, the energy required for amplification is provided by a thermal gradient rather than an externally applied electrical bias.

Here, we introduce a zero-bias superconducting voltage amplifier that takes advantage of the bipolar thermoelectric effect~\cite{marchegiani2020nonlinear, germanese2022bipolar, germanese2023phase, Antola_2024, antola2025quantumbipolarthermoelectricity} in asymmetric superconductor-insulator-superconductor tunnel junctions (SIS). The device operates without any applied DC bias, and the amplification mechanism converts low-frequency input voltage modulations into AC output signals, while ensuring minimal power dissipation at millikelvin temperatures. The underlying operating principle is to maintain one electrode at an elevated electronic temperature ($\sim 1\,\mathrm{K}$), while the other remains thermalized in the $20\,\mathrm{mK}$ cryogenic bath. Under these nonequilibrium thermal conditions, the combined effect of the asymmetric superconducting density of states and thermal broadening enhances the quasiparticle tunneling rate across the junction. For voltages approaching the difference between the superconducting energy gaps of the two electrodes, this asymmetry generates a current that flows opposite to the applied voltage, thereby giving rise to a negative absolute and differential resistance region ($\mathrm{d}V/\mathrm{d}I < 0$). When placed in series with a load resistor, the junction provides voltage amplification in a voltage-divider configuration without an external bias source.

The experimental feasibility of the proposed device is substantiated by previous work: the superconducting bipolar thermoelectric  effect has been demonstrated in analogous asymmetric junctions~\cite{germanese2022bipolar}, with an estimated thermal load not exceeding $\sim 1$~nW, which remains well within the cooling power of conventional dilution refrigerators. In addition, the constituent materials and fabrication processes are based on $\text{Al}$, $\text{Al-Cu}$, and $\text{AlO}_x$, all of which are standard in superconducting circuit technology~\cite{giazotto2006opportunities}. Numerical simulations indicate a voltage gain of 20 dB with microvolt input saturation, total harmonic distortion below $-50$ dB for input amplitudes up to $1~\mu$V, input-referred noise below 1 nV/$\sqrt{\mathrm{Hz}}$, and an operational bandwidth extending from near DC up to approximately 180 MHz. The device is designed to operate at bath temperatures in the range 20-250 mK and is fully compatible with state-of-the-art aluminum-based superconducting circuit fabrication processes. 
\section{Device Model} \label{sec:device model}

The device comprises an asymmetric superconductor–insulator–superconductor (SIS) tunnel junction connected in series with a load resistor $R_L$, as illustrated in Fig.~\ref{fig:fig1}(a). The junction is realized by combining a pure aluminum electrode ($\Delta_2 = 0.20$~meV, $T_{c,2} = 1.3$~K) with an aluminum–copper proximity bilayer ($\Delta_1 = 0.10$~meV, $T_{c,1} = 0.65$~K), separated by an $\text{AlO}_x$ tunnel barrier. This architecture yields a gap ratio $\Delta_1(0)/\Delta_2(0) = 0.5$, which is in the vicinity of the theoretically predicted optimal regime for the bipolar thermoelectric effect and has been subsequently experimentally validated in asymmetric superconducting tunnel junctions~\cite{marchegiani2020nonlinear,germanese2022bipolar,germanese2023phase}. 
Comparable performance is achieved over a broader range of gap ratios, $\Delta_1(0)/\Delta_2(0) \sim 0.4$–$0.8$, provided that the device parameters are appropriately rescaled. In the following, we adopt $\Delta_1(0)/\Delta_2(0) = 0.5$ as a representative example.

The electrode with a reduced superconducting gap is implemented using an Al–Cu proximity bilayer, whose superconducting critical temperature and energy gap are tunable by appropriately choosing the thicknesses of the individual layers. Such bilayer systems have been extensively investigated and experimentally characterized, demonstrating reliable and controllable suppression of the superconducting gap within aluminum-based nanofabrication platforms~\cite{PhysRevApplied.14.034055,germanese2022bipolar,germanese2023phase,de2024transition}. In operation, the aluminum electrode is heated to $T_H \sim 1$~K, while the bilayer is thermally anchored to the bath at temperature $T_B$. The input signal $V_{in}$ is applied to the series combination of the junction and the load, and the output voltage $V_{out}$ is measured across the junction. DC bias is not applied.

Thermal gradients of comparable magnitude have already been experimentally demonstrated in superconducting tunnel junctions employing quasiparticle injection heaters \cite{germanese2022bipolar,germanese2023phase}. In those devices, the electronic temperature of one superconducting electrode was raised to approximately 1 K, while the counter-electrode remained near the bath temperature. Although the junction resistance in the present design is lower, thus improving heat transport through the tunnel barrier, the associated additional heat leakage can be mitigated by integrating normal-metal quasiparticle traps or \textit{cooling fins} on top of the cold junction electrode, a strategy routinely implemented in superconducting microcoolers \cite{giazotto2006opportunities,pekola2004limitations,tirelli2008manipulation,quaranta2011cooling,PhysRevApplied.23.014046} and coherent caloritronic devices \cite{fornieri2017towards,fornieri20170,timossi2018phase}.

Zero-voltage bias constitutes the intrinsic operating point of the device. This obviates the need for an external DC biasing network and the corresponding power dissipation at millikelvin temperatures: the amplification process is driven exclusively by the imposed thermal gradient, with a power budget constrained to the nanowatt scale at most \cite{germanese2022bipolar,germanese2023phase} due to the heaters necessary to increase the electronic temperature of the hot electrode. Because the $I$–$V$ characteristic (Eq.~\ref{eq:tunnel}) is an odd function of the applied voltage, the zero-bias operating point is located at the center of the negative-resistance region. This symmetry guaranties the cancelation of all even-order harmonic distortion components, thereby enhancing linearity and maximizing the allowable symmetric input excursion before the onset of compression. As a result, zero-voltage bias yields an optimal compromise between gain and dynamic range and is adopted as the default operating condition for all analyzes reported in this work.

The quasiparticle tunneling current through an SIS junction is given by~\cite{giazotto2006opportunities, marchegiani2020nonlinear}
\begin{align}
I_T(V_T, T_B, T_H) &= \frac{1}{eR_T} \int_{-\infty}^{+\infty} N_1(\epsilon, T_B)\, N_2(\epsilon - eV_T, T_H)\nonumber \\
    &\quad \times [f(\epsilon - eV_T, T_H) - f(\epsilon, T_B)]\, \mathrm{d}\epsilon,
\label{eq:tunnel}
\end{align}
where $R_T$ is the tunnel resistance in the normal-state, $N_{1,2}(E, T)$ are the normalized Bardeen-Cooper-Schrieffer (BCS) superconducting densities of states (DOS) of the two electrodes, and $f(E, T) = [1 + \exp(E/k_B T)]^{-1}$ is the Fermi-Dirac distribution
at temperature $T$. The normalized BCS density of states is given by
\begin{equation}
N_{i}(\epsilon, T) = \left|\mathrm{Re}\left(\frac{\epsilon + i\Gamma_i}
{\sqrt{(\epsilon + i\Gamma_i)^2 - \Delta_i^2(T)}}\right)\right|,
\label{eq:bcs}
\end{equation}
where $\Delta_i$ is the energy gap of electrode $i$ and $\Gamma_i$ is a phenomenological Dynes broadening parameter accounting for finite quasiparticle states within the gap.
The temperature dependence of the gap is obtained from the approximated BCS self-consistency equation~\cite{gross_anomalous_1986}
\begin{align}
\Delta_i(T) \simeq \Delta_i(0) \tanh{\left(1.74\sqrt{\frac{T_{c,i}}{T} - 1}\right)},
\end{align}
where $T_{c,1}=0.65$ K and $T_{c,2}=1.3$ K. Both $\Delta_1(T_B)$ and $\Delta_2(T_H)$ are evaluated at their respective electrode electronic temperatures.

For a junction area of $1~\mu\text{m}^2$, we assume a tunnel resistance $R_T = 100~\Omega$ and capacitance $C_T = 50~\text{fF}$. Therefore, the specific values are $\rho_T = 100~\Omega\,\mu\text{m}^2$ and $c_T = 50~\text{fF}/\mu\text{m}^2$, typical values for relatively high-transparency SIS junctions \cite{10.1063/1.2357915}. The Dynes broadening parameters are set to $\Gamma_{1,2} = 0.075\Delta_{{1,2}}(0)$ to ensure stable amplifier operation. By smoothing the gap-edge DOS divergence, this level of broadening prevents the $I$-$V$ characteristic from developing sharp peaks that could lead to multiple load-line intersections. Such intersections would create stable fixed points in regions of positive differential resistance, trapping the operating point, and quenching the NDR. 
The chosen $\Gamma$ ensures a monotonic $I$-$V$ slope throughout the NDR region, guaranteeing a unique solution for $R_L < |r_d|$. Experimentally, these broadening levels and the suppression of the Josephson current (neglected here) can be achieved in high-transparency tunnel junctions under weak magnetic fields \cite{germanese2022bipolar,germanese2023phase} or at high temperatures~\cite{Milliken2004_NbSIS, PhysRevLett.105.026803}.

\begin{figure*}[t!]
\includegraphics[width=\linewidth]{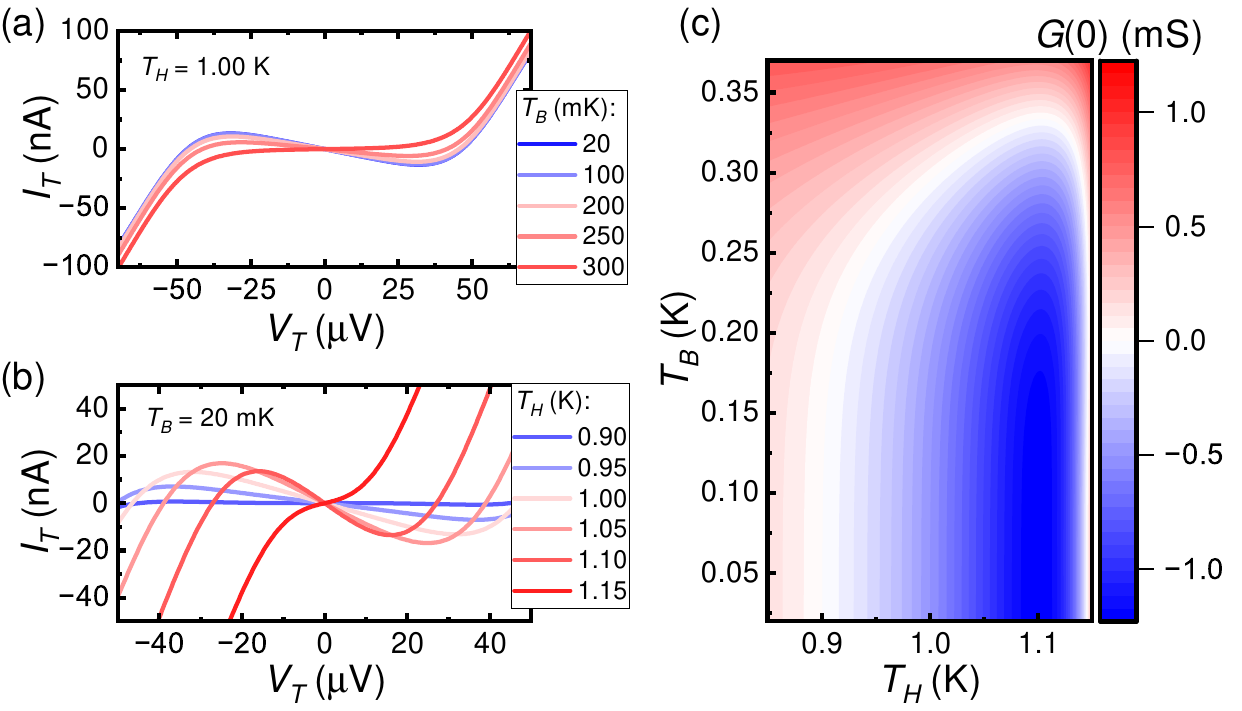}
\caption{\label{fig:fig2} Current–voltage characteristics and differential conductance as functions of temperature.  
(a) Current–voltage ($I$–$V$) characteristics as a function of the bath temperature $T_B$ at fixed hot-electrode temperature $T_H = 1$~K. For $T_B < 0.3$~K, the $I$–$V$ curves display a region of negative differential conductance extending over the bias interval $|V_T| \lesssim \Delta_2(T_H) - \Delta_1(T_B)$. This region vanishes for $T_B \gtrsim 0.3$~K, in correspondence with the suppression of the bipolar thermoelectric effect.  
(b) $I$–$V$ characteristics as a function of the hot-electrode temperature $T_H$ at fixed $T_B = 20$~mK. The occurrence of negative differential conductance is highly sensitive to $T_H$: for $T_H \gtrsim 1.1$~K, the superconducting gap $\Delta_2(T_H)$ is substantially suppressed, whereas for $T_H \lesssim 0.9$~K, thermal broadening is insufficient to significantly populate quasiparticle states above the gap edge. Optimal device performance is achieved at $T_H \simeq 1$~K with $T_B = 20$~mK.  
(c) Zero-bias conductance map $G(0)$ as a function of $T_H$ and $T_B$. The blue region ($G(0) < 0$) indicates the amplification regime associated with negative differential conductance, whereas the red region corresponds to the positive-conductance regime, in which the thermoelectric response is effectively quenched.}
\end{figure*}

The bipolar thermoelectric effect, schematically illustrated in Fig.~\ref{fig:fig1}(b), is based on an asymmetry between the superconducting energy gaps ($\Delta_1(0) < \Delta_2(0)$) in combination with a strong temperature gradient ($T_H \gg T_B$), such that $\Delta_2(T_H) > \Delta_1(T_B)$. For small positive bias voltages ($0 < eV \lesssim \Delta_2(T_H) - \Delta_1(T_B)$), thermal broadening in the hot electrode populates the quasiparticle states above $\Delta_2$ in a manner that brings them into energetic alignment with the sharp DOS singularity at the gap edge of the cold electrode. This alignment permits quasiparticles in these thermally excited states to tunnel resonantly into the DOS peak, thereby generating a net charge current that flows opposite to the applied bias. Due to particle–hole symmetry, this counter-propagating current is preserved upon reversal of the bias polarity, resulting in $IV < 0$ for both positive and negative voltages. The subsequent inversion of the current–voltage characteristic produces a negative absolute resistance ($V/I < 0$), as well as a negative differential resistance ($r_d = (\text{d}I/\text{d}V)^{-1} < 0$), in the vicinity of zero bias.

The amplification mechanism is schematically illustrated in Fig.~\ref{fig:fig1}(c). The $I$–$V$ characteristic of the junction exhibits a region of negative differential resistance in the vicinity of zero bias. The load resistor $R_L$ defines a load line whose slope intersects the $I$–$V$ characteristic at the quiescent operating point at zero bias. A small superimposed AC input signal $V_{in}$ induces perturbations of this operating point along the load line, thus generating a correspondingly modulated and amplified output voltage $V_{out}$ across the junction.
Since the circuit operates as a voltage divider, the voltage gain can be written as
\begin{equation}
G_V = \frac{|r_d|}{|r_d| - R_L},
\label{eq:gain}
\end{equation}
where $r_d = (\text{d}I/\text{d}V)^{-1} < 0$ denotes the differential resistance, evaluated at zero bias. The gain exceeds unity when the condition $|r_d| > R_L$ is satisfied, in which case the energy required for amplification is provided solely by the thermal gradient, rather than by an external DC bias source.

The dependence of the junction $I$–$V$ characteristics and the zero-bias conductance $G(0)$ on the electrode temperatures is presented in Fig.~\ref{fig:fig2}. Figure~\ref{fig:fig2}(a) displays the probe current $I_T$ as a function of the probe voltage $V_T$ at a fixed hot-electrode temperature $T_H = 1$~K, for bath temperatures $T_B$ that range from $20$~mK to $300$~mK. A region of negative differential resistance (NDR) is observed for $T_B \lesssim 300$~mK, with an associated voltage interval $|V_T| \lesssim |\Delta_2(T_H) - \Delta_1(T_B)|$ that monotonically decreases as $T_B$ increases. 
In contrast, at fixed $T_B = 20$~mK, the NDR exhibits a pronounced dependence on $T_H$, as shown in Fig.~\ref{fig:fig2}(b). For $T_H \lesssim 0.9$~K, thermal broadening of the quasiparticle energy distribution is insufficient to give rise to a well-defined NDR region, whereas for $T_H \gtrsim 1.15$~K, the superconducting gap $\Delta_2(T_H)$ is substantially suppressed, thus quenching the effect. This thermal balance is summarized in Fig.~\ref{fig:fig2}(c), where the space of operational parameters corresponding to $G(0) < 0$ is delineated by the temperature constraints required to sustain both an adequate quasiparticle population and the inequality $\Delta_2(T_H) > \Delta_1(T_B)$. On this basis, $T_H \simeq 1$~K and $T_B = 20$~mK are identified as the optimal operating conditions for the results reported in this work.

The frequency response and dynamical behavior of the amplifier are obtained by numerically integrating the full nonlinear circuit equation
\begin{align}
C_T \dot{V}_T(t) = \frac{V_{in}(t) - V_T(t)}{R_L} - I_T(V_T, T_B, T_H),
\label{eq:ode}
\end{align}
where $V_{in}(t) = V_{0}\sin(\omega t)$ denotes a purely AC input signal with zero DC offset, $I_T(V_T, T_B, T_H)$ represents the complete nonlinear tunneling current as defined in Eq.~\ref{eq:tunnel}, and $V_T(t)$ is the time-dependent output voltage across the junction. The voltage gain and harmonic spectrum are subsequently determined from the steady-state temporal evolution of $V_T(t)$. This approach intrinsically captures both saturation phenomena and the $RC$ roll-off associated with $C_T$, without resorting to small-signal linearization, and is employed to obtain all results presented in Fig.~\ref{fig:fig3} and Fig.~\ref{fig:fig4}.

The input-referred voltage noise, $\sqrt{S_V^{\mathrm{in}}}$, is evaluated at the zero-bias operating point. In this analysis, two independent current-noise contributions are considered: the shot noise generated by the tunnel junction and the Johnson–Nyquist noise associated with the load resistor. 
The shot noise of the tunnel junction is computed using the full bidirectional tunneling rates \cite{giazotto2006opportunities,golubevNonequilibriumTheoryHotelectron2001, cz4n-rh4r} as
\begin{align}
S_I^{\mathrm{shot}}(V_T) &= \frac{2}{R_T} \int_{-\infty}^{+\infty} N_1(\epsilon, T_B)\,
N_2(\epsilon - eV_T, T_H)  \nonumber \\ 
    &\quad \times \bigl[f(\epsilon - eV_T, T_H)\bigl(1 - f(\epsilon, T_B)\bigr) \nonumber \\ 
    &\quad + f(\epsilon, T_B)\bigl(1 - f(\epsilon - eV_T, T_H)\bigr)\bigr]\, \mathrm{d}\epsilon,
\label{eq:shotnoise}
\end{align}
which explicitly incorporates contributions from both forward and backward tunneling processes. The Johnson–Nyquist noise of the load resistor is given by $S_I^{\mathrm{load}} = 4k_B T_L / R_L$, where $T_L$ denotes the electronic temperature of the resistor $R_L$. The value of $T_L$ depends on the physical location of $R_L$ within the cryogenic environment, as will be discussed in detail below.
The total current-noise spectral density, $S_I^{\mathrm{tot}} = S_I^{\mathrm{shot}} + S_I^{\mathrm{load}}$, is converted to an output-voltage noise spectral density through the circuit admittance according to
\begin{equation}
S_V(\omega) = \frac{S_I^{\mathrm{tot}}}{g_{\mathrm{eff}}^2 + (\omega C_T)^2},
\label{eq:svout}
\end{equation}
where $g_{\mathrm{eff}} = 1/R_L + g_d$ is the effective conductance at the operating point, and $g_d = \mathrm{d}I_T/\mathrm{d}V_T\big|_{V=0} < 0$ is the differential conductance of the junction. The input voltage-noise spectral density is then obtained by normalizing with respect to the squared voltage gain, resulting in $S_V^{\mathrm{in}}(\omega) = S_V(\omega)/G_V^2$. 
Two additional noise contributions are neglected in this analysis: the heat-current fluctuations and the heat–charge-current cross-correlator associated with the tunnel junction~\cite{giazotto2006opportunities,golubevNonequilibriumTheoryHotelectron2001, cz4n-rh4r}. This approximation is justified under the assumption that both electrodes remain in local thermal equilibrium at their respective temperatures $T_H$ and $T_B$.
This assumption is equivalent to requiring that the thermal capacitances of the electrodes are sufficiently large such that temperature fluctuations can be neglected and the corresponding thermal susceptibilities, $\chi^{th}_{1,2}(\omega)=\frac{1}{g^{\mathrm{th}}_{1,2}+i\omega C^{\mathrm{th}}_{1,2}}$, effectively vanish. Here, $g^{\mathrm{th}}_{1,2}$ denotes the thermal conductance and $C^{\mathrm{th}}_{1,2}$ the thermal capacitance of each electrode, respectively.

\section{Results and Discussion} \label{sec:results}

\begin{figure}[t!]
\includegraphics[width=\linewidth]{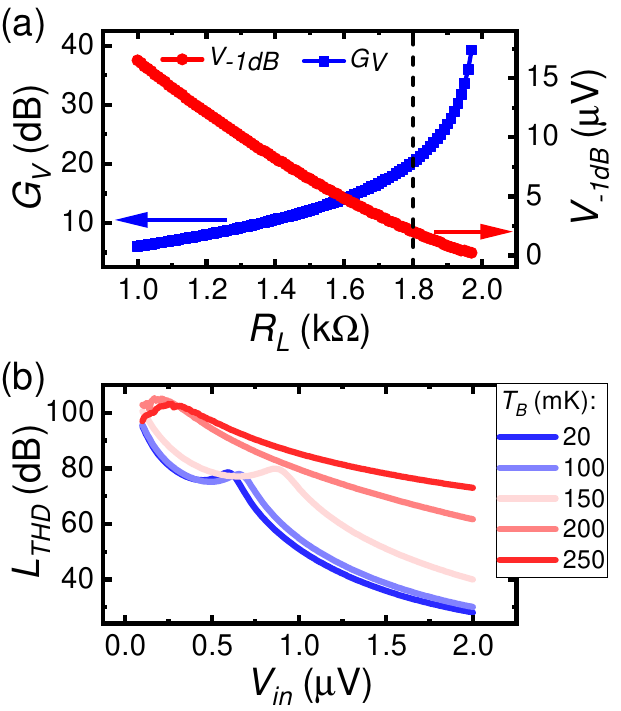}
\caption{\label{fig:fig3} 
Figures of merit: gain, saturation, and linearity.  
(a) Voltage gain $G_V$ (blue) and 1~dB compression point $V_{-1\text{dB}}$ (red) as functions of load resistance $R_L$ at $T_B = 20$~mK and $T_H = 1$~K. The gain is given by $G_V = |r_d|/(|r_d| - R_L)$, where $r_d = (dI/dV)^{-1} \simeq -2000~\Omega$. As $R_L$ approaches $|r_d|$, the gain diverges, whereas the 1~dB compression point tends to zero. The vertical dashed line indicates the chosen operating point $R_L = 1800~\Omega$, which yields $G_V \simeq 20$~dB and $V_{-1\text{dB}} \simeq 2~\mu$V.  
(b) Linearity metric $L_{\text{THD}}$ as a function of input amplitude $V_{\text{in}}$ for $R_L = 1800~\Omega$ at several bath temperatures. The quantity $L_{\text{THD}}$ is defined as $-20\log_{10}\!\bigl(\sqrt{\sum_{i=2}^{\infty} A_i^2/A_1}\bigr)$, where $A_i$ denote the output harmonic amplitudes. Linearity deteriorates with increasing drive amplitude. In contrast, increasing bath temperature improves linearity due to the reduction of gain, which suppresses nonlinear distortion. The non-monotonic features originate from a cancellation of the third-order harmonic contribution to the total harmonic distortion (see Appendix \ref{sec:appendix1}).}
\end{figure}

The performance of the amplifier is quantitatively specified in terms of its voltage gain $G_V$, input saturation level $V_{-1\mathrm{dB}}$, linearity expressed by the total harmonic distortion $L_{THD}$, frequency response, and input-referred voltage noise spectral density $\sqrt{S_V^{\mathrm{in}}}$.

The trade-off between voltage gain and dynamic range is illustrated in Fig.~\ref{fig:fig3}(a), which presents $G_V$ and the 1~dB compression point $V_{-1\mathrm{dB}}$ as functions of the load resistance $R_L$ at $T_B = 20$~mK and $T_H = 1$~K. As $R_L$ approaches the magnitude of the differential resistance, $|r_d| \simeq 2000$~$\Omega$, the gain increases sharply in accordance with Eq.~\ref{eq:gain}, whereas $V_{-1\mathrm{dB}}$ decreases toward zero. In the limit $R_L \to |r_d|$, the gain formally diverges while the linear input range vanishes. The selected operating point, $R_L = 1800$~$\Omega$ (indicated by the vertical dashed line), represents a compromise, yielding a voltage gain of approximately 20~dB and a 1~dB compression point of approximately 2~$\mu$V. This operating point is adopted throughout this work.

The linearity of the amplifier is characterized by the total harmonic distortion (THD), defined as
\begin{equation}
L_{THD} = -20\log_{10}\left(\frac{\sqrt{\sum_{i=2}^{\infty} A_i^2}}{A_1}\right),
\label{eq:thd}
\end{equation}
where $A_i$ denotes the amplitude of the $i$-th output harmonic, obtained from the fast Fourier transform (FFT) of the output signal. Larger values of $L_{THD}$ correspond to lower distortion. Figure~\ref{fig:fig3}(b) presents $L_{THD}$ as a function of the input amplitude $V_{in}$ for $R_L = 1800$~$\Omega$, a hot-electron temperature $T_H = 1$~K, and bath temperatures $T_B$ in the range $20$~mK–$250$~mK. In the low-amplitude regime, $L_{THD}$ exceeds $100$~dB, indicating that distortion is negligibly small under small-signal conditions. As $V_{in}$ is increased, $L_{THD}$ decreases, reflecting the onset of saturation and the consequent degradation of linearity. An increase in $T_B$ improves linearity, as the reduction of the thermoelectric response lowers the gain and consequently suppresses nonlinear distortion, although at the expense of reduced amplification. The non-monotonic features observed at intermediate input amplitudes are attributed to the crossover between third- and fifth-order harmonic contributions, as detailed in Appendix~\ref{sec:appendix1}.

\begin{figure}[t!]
\includegraphics[width=\linewidth]{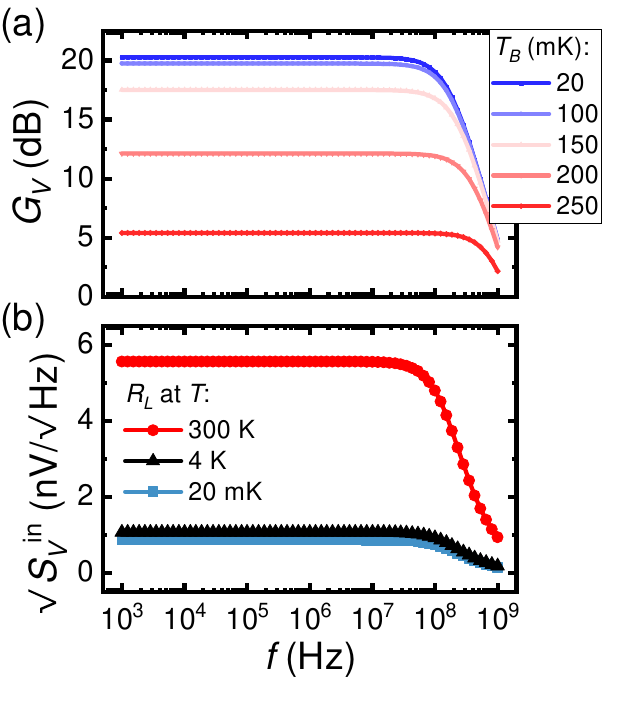}
\caption{\label{fig:fig4} Frequency response: gain and input-referred voltage noise.  
(a) Voltage gain $G_V$ as a function of frequency $f$ for various bath temperatures $T_B$ at fixed $T_H = 1$~K and $R_L = 1800~\Omega$. The response is broadband, starting near DC, with a $-3$~dB cutoff at around $180$~MHz, set by the junction RC time constant. Increasing $T_B$ suppresses the thermoelectric response, leading to an increase in $|r_d|$ and a corresponding reduction in gain.  
(b) Input-referred voltage noise spectral density $\sqrt{S_V^{\mathrm{in}}}$ as a function of frequency for different load resistor temperatures. For an on-chip resistor at $T_B = 20$~mK, the noise is predominantly determined by shot noise originating from the hot electrode, resulting in $\sqrt{S_V^{\mathrm{in}}} = 0.86$~nV/$\sqrt{\text{Hz}}$. As the load resistor temperature is increased, its Johnson noise contribution eventually becomes the dominant component of the total noise floor. The noise spectrum is essentially frequency-independent from near-DC up to the amplifier bandwidth, beyond which it rolls off.}
\end{figure}

The frequency response of the amplifier is presented in Fig.~\ref{fig:fig4}(a), where the voltage gain $G_V$ is plotted as a function of frequency $f$ for several values of bath temperature $T_B$, at fixed hot temperature $T_H = 1$~K and load resistance $R_L = 1800$~$\Omega$. 
The gain remains approximately constant from near-DC up to a $-3$~dB cutoff frequency of about $180$~MHz, which is determined by the $RC$ time constant of the tunnel junction. 
The gain decreases monotonically with increasing $T_B$, in agreement with the previously discussed temperature dependence of the differential resistance $r_d$: as the bath temperature increases, the bipolar thermoelectric effect weakens, resulting in an increase of $|r_d|$ and a corresponding reduction of $G_V$ towards unity. 
This cutoff can be derived analytically from the equivalent circuit model. In particular, the junction capacitance $C_T$ is periodically charged and discharged through the parallel combination of $R_L$ and $|r_d|$, leading to an effective time constant
$\tau = \frac{C_T R_L |r_d|}{|r_d| - R_L}$
and an associated $-3$~dB cutoff frequency
\begin{equation}
f_{-3\text{dB}} = \frac{1}{2\pi\tau} = \frac{|r_d| - R_L}{2\pi C_T R_L |r_d|}.
\label{eq:cutoff}
\end{equation}
For $R_L = 1800~\Omega$, $|r_d| \simeq 2000~\Omega$, and $C_T = 50$~fF, we obtain $f_{-3\text{ dB}} \simeq 180$~MHz, in good agreement with the simulated cutoff frequency. It is important to note that in the limit $R_L \to |r_d|$, where the small-signal gain formally diverges, the bandwidth simultaneously collapses, i.e., $f_{-3\text{ dB}} \to 0$. Consequently, the condition that maximizes gain inherently minimizes bandwidth, such that the gain–bandwidth product
$G_V f_{-3\text{ dB}} = \frac{1}{2\pi C_T R_L} \simeq 1.8~\text{GHz}$
remains constant and is determined solely by the junction capacitance and the load resistance.

The input-referred voltage noise spectral density, $\sqrt{S_V^{in}}$, is presented in Fig.~\ref{fig:fig4}(b) for three different configurations of the load resistor operated at different temperatures. When $R_L$ is integrated on-chip and thermally anchored to the mixing chamber at $T = 20$~mK, the Nyquist–Johnson noise associated with $R_L = 1800$~$\Omega$ is negligible compared to the shot noise generated by the heated electrode at $T_H = 1$~K. Under these conditions, the input-referred voltage noise remains approximately constant at $\sqrt{S_V^{in}} = 0.86$~nV/$\sqrt{\text{Hz}}$ from near-DC up to the bandwidth limit. 
When $R_L$ is instead mounted at the $4$~K temperature stage, its Johnson noise becomes comparable to the junction shot noise, leading to an increase in the total input-referred noise floor to about $1.1$~nV/$\sqrt{\text{Hz}}$. These noise levels are consistent with those reported for state-of-the-art cryogenic semiconductor amplifiers~\cite{Ivanov2011, zavjalovCryogenicDifferentialAmplifier2019}. At room temperature ($T = 300$~K), the thermal noise of $R_L$ dominates the noise budget, yielding $\sqrt{S_V^{in}} = 5.6$~nV/$\sqrt{\text{Hz}}$. 
Across all configurations, the noise spectral density remains approximately frequency-independent from near DC and exhibits a roll-off at frequencies exceeding the amplifier $RC$ cutoff. 
Taken together, these results indicate that positioning $R_L$ in the millikelvin or $4$~K stage is essential to maintain near-optimal noise performance. 
We note that replacing, for example, $R_L$ with a cryo-CMOS transistor operated at the $4$~K stage would enable in situ tuning of the load impedance, thereby allowing real-time optimization of the gain–saturation and gain–bandwidth trade-off while preserving comparable noise performance.
\section{Conclusion} \label{sec:conclusion}

In conclusion, we have demonstrated a zero-bias superconducting voltage amplifier that operates on thermally induced negative differential resistance in asymmetric SIS junctions. By harnessing the bipolar thermoelectric effect in a junction characterized by a zero-temperature gap ratio $\Delta_1(0)/\Delta_2(0) = 0.5$ and subject to a temperature gradient from $T_H \simeq 1$~K to $T_B \simeq 20$~mK, the device realizes voltage amplification without the need for DC bias power. The structure, comprising aluminum and aluminum–copper electrodes separated by an $\text{AlO}_x$ tunnel barrier, is fully compatible with standard aluminum-based superconducting fabrication processes.

Numerical simulations indicate a voltage gain of 20~dB with a 1~dB compression point occurring at an input amplitude of 2~$\mu$V, while maintaining total harmonic distortion below $-50$~dB for sub-microvolt input levels. When the load resistor is thermally anchored either on-chip or at the 4~K stage, the resulting input-referred noise spectral density is approximately $1~\text{nV}/\sqrt{\text{Hz}}$, which is comparable to that of state-of-the-art cryogenic semiconductor amplifiers. In addition, the device demonstrates a broadband frequency response extending from near-DC, with a $-3$~dB cutoff frequency around 180~MHz.

These characteristics render the device well-suited for cryogenic detector readout and for low-noise measurements of mesoscopic quantum systems. More broadly, bias-free superconducting voltage amplification at millikelvin temperatures has the potential to complement cryogenic semiconductor amplifiers operating at 4 K, thereby enabling more compact and highly integrated cryogenic instrumentation. Future research will focus on experimental verification and quantitative performance benchmarking.

\section*{Acknowledgements} \label{sec:acknowledgements}

We acknowledge F. Antola and A. Braggio for their insightful and stimulating discussions.  
The authors gratefully acknowledge partial financial support from the PNRR MUR project PE0000023-NQSTI.

\section*{Data availability}

Data supporting the findings of this article are openly available \cite{trupiano_2026_19070947}.

\appendix
\section{Linearity analysis} \label{sec:appendix1}

The non-monotonic features in Fig.~\ref{fig:fig3}(b) arise from a cancelation of the third harmonic contribution to the total harmonic distortion at a specific input amplitude value. To understand this, we expand the quasi-static transfer function $V_{out}(V_{in})$ in a Taylor series around the zero-bias operating point:
\begin{equation}
V_{out}(V_{in}) = a_1 V_{in} + a_3 V_{in}^3 +
a_5 V_{in}^5 + \dots
\label{eq:taylor}
\end{equation}

Only odd-order terms appear, as required by the odd symmetry of the $I$-$V$
characteristic at zero bias. For a sinusoidal input $V_{in}(t) = A\sin(\omega t)$, trigonometric identities yield the amplitudes of the fundamental and odd harmonics at the output:
\begin{align}
H_1(A) &= \left|a_1 A + \frac{3}{4}a_3 A^3 + \frac{5}{8}a_5 A^5\right|,
\label{eq:H1}\\
H_3(A) &= \left|\frac{1}{4}a_3 A^3 + \frac{5}{16}a_5 A^5\right|,
\label{eq:H3}\\
H_5(A) &= \left|\frac{1}{16}a_5 A^5\right|.
\label{eq:H5}
\end{align}
At small amplitudes, $H_3 \propto A^3$ dominates the distortion and $L_{THD}$ decreases monotonically as $A$ increases, as expected for a weakly nonlinear amplifier. The behavior becomes richer at intermediate amplitudes due to the sign structure of the Taylor coefficients. For the transfer function $V_{out}(V_{in})$ of this device, $a_3 < 0$ and $a_5 > 0$ As a result, the two contributions to $H_3$ in Eq.~\ref{eq:H3} have opposite signs and cancel at the amplitude
\begin{equation}
A_{peak} = \sqrt{-\frac{4a_3}{5a_5}},
\label{eq:apeak}
\end{equation}
where $H_3$ reaches a local minimum. At this specific amplitude the third harmonic is suppressed, and the total distortion is temporarily dominated by $H_5 \propto A^5$, which is smaller than $H_3$ would have been without cancelation. This produces the local maximum in $L_{THD}$ visible in Fig.~\ref{fig:fig3}(b). Beyond $A_{peak}$, both $H_3$ and $H_5$ grow with increasing amplitude. The non-monotonic  jump is therefore a direct consequence of the opposite signs of $a_3$ and $a_5$, which in turn reflect the specific shape of the NDR region of the $I$-$V$ characteristic.

The validity of this quasi-static analysis is confirmed by two complementary approaches. First, a linear regression of the transfer function $V_{out}(V_{in})$ over a sliding amplitude window shows that the coefficient of determination $R^2$ reaches a local maximum near $A_{peak}$, consistent with the temporary restoration of linearity due to the cancelation of $H_3$. Second, the harmonic amplitudes $H_1$, $H_3$, and $H_5$ are extracted directly from the Fast Fourier Transform (FFT) of the full nonlinear ODE solution of Eq.~\ref{eq:ode}, confirming both the amplitude dependence and the location of the cancelation peak $A_{peak}$.

\bibliography{bibliography}

\end{document}